\RequirePackage{fix-cm}
\documentclass[smallextended]{svjour3}
\smartqed
\journalname{J Supercond Nov Magn}
\begin{document}

\title{Cuprates - An Overview}

\author{M. R. Norman}

\institute{Materials Science Division, Argonne National Laboratory, Argonne, IL  60439, USA\\
              \email{norman@anl.gov}}

\date{\today}

\maketitle

\begin{abstract}
A brief overview is given of the problem of high temperature superconductivity in the cuprates,
with an emphasis on theoretical ideas.
\end{abstract}

What does it mean to say that we have ``solved" the cuprate problem?  To understand
this point, let us consider the classic case of $^3$He.  In the late 1960s, it had been predicted
that $^3$He would be a p-wave superfluid mediated by exchange of spin fluctuations \cite{layzer}.
Indeed, several years later, this was discovered to be the case, although two superfluid phases
were found, which were denoted as A and B.  Shortly thereafter, Anderson and Brinkman
were able to show that the existence of the A phase could be attributed to the feedback 
on the spectrum of spin
fluctuations from the energy gap due to pairing \cite{brinkman}.  At that point, one might have decided
to declare victory and move on.  But as we now know, the situation is not nearly so straightforward as that.
Because of the simplicity of this system (a single spherical Fermi surface with a small ratio
of $T_c$ to $E_F$), one can to a good approximation determine the potential for Cooper pairing
from knowledge of the Landau parameters of the normal state.  When one does this,
one finds that a variety of interactions contribute to the pairing potential, including density, spin,
and transverse current fluctuations \cite{leggett}.  This is also reflected in microscopic theories.
At the time heavy fermion superconductors were discovered and theorists began to
migrate out of the $^3$He field in the early 1980s to tackle that problem, a number of theories existed
which stressed different parts of the interaction.  To this date, no resolution has occurred to pick out
one of these over the others.  So, can we truly say we have `solved' the problem of $^3$He?

When considering cuprates, other complicating factors enter which do not occur in $^3$He.  Cuprates
have a lattice, thus breaking translational symmetry.  Different types of atoms are involved, with multiple orbitals
per atom.  When addressing band theory results, a minimum of four orbitals is needed to properly describe
the low energy sector.  Even then, it has been argued that by concentrating on the low energy subspace
(a single anti-bonding band of hybridized Cu 3$d_{x^2-y^2}$ and O 2$p_x$ and 2$p_y$ states), 
one is throwing out the baby with the
bathwater.  For instance, the theory of Varma \cite{varma}, which advocates the existence of orbital
currents flowing within the primitive unit cell, specifically requires considering at least three energy bands
in the microscopic theory.  Add to this the importance of lattice interactions (known to be particularly 
important because of the presence of polaronic-like effects) and the existence of spin-orbit
coupling, and the problem becomes complex, especially due to the large Coulomb interaction associated
with the copper sites, as realized early on by Anderson \cite{pwa}.

In fact, the wealth of theoretical ideas that have emerged when thinking about the cuprate problem has
been a boon to condensed matter physics.  No matter its pros and cons, the resonating valence bond
(RVB) theory of Anderson \cite{pwa} has had a profound influence on physics, leading to a revival in 
the study of quantum spin liquids.  The prediction of stripe formation in cuprates and their
subsequent observation has led to a revolution in techniques designed to probe such states, including
neutron scattering and scanning tunneling microscopy \cite{RMP-Steve}.
The latter (STM) has really acted to highlight the importance
of inhomogeneity in the physics of the cuprates, particularly for underdoped compounds \cite{STM}.
And the existence of non-Fermi liquid like behavior as predicted by a number of theories and first revealed 
by transport measurements, has led to astonishing advances in such techniques as angle resolved
photoemission that can directly probe the single-particle spectral function \cite{RMP-ZX}.

But perhaps the most important development has been to focus our attention on the problem of strong
correlations.  For classic superconductors, it only took a few years between the advent of the weak coupling
theory of Bardeen, Cooper and Schrieffer \cite{bcs} to the development of a full strong coupling
theory \cite{scalapino}.  This amazing development can be attributed to something known as Migdal's
theorem.  In classic superconductors, pairing is mediated by exchange of phonons.  Since 
the energy of the 
phonons is much smaller than the Fermi energy, one has a controlled perturbation
expansion in the small ratio $\hbar\omega_D/E_F$.  In fact, for most properties, one can stop at lowest
order when calculating the self-energy (unless polaronic effects are important).  Therefore, the only
diagram series that has to be summed is the particle-particle ladder one due to repeated scattering of
electrons by phonons which defines the pairing instability \cite{AGD}.  Moreover, this small ratio also
has the beneficial effect of reducing the repulsive contribution coming from the Coulomb 
interaction when scaling it from $E_F$ down to $\hbar\omega_D$ \cite{morel}.  But the price one pays
is that this same `retardation' effect limits $T_c$ \cite{cohen}.

For pairing due to electron-electron interactions, though, all bets are off.  The collective degrees of freedom
such as spin fluctuations are composed of the very same electrons one is trying to pair  (as opposed
to the electron-ion case, where electrons and phonons can be treated to a good approximation as separate
entities).  One can certainly try to invoke a modified Migdal theorem by exploiting the small ratio
of collective energies such as $\hbar\omega_{sf}$ to $E_F$, but there is no rigorous
grounds for doing so \cite{chubukov}.  Back in the early days of $^3$He, it was realized that vertex corrections 
could be as large as the leading order term in a perturbation expansion \cite{levin}.

One can attempt to bypass these difficulties by a variety of techniques.  For instance, one can try to use
dressed Greens functions in the expansion (the so-called fluctuation exchange or FLEX 
approximation) \cite{flex}, but this leads
to the suppression of correlation gaps \cite{vilk} in clear contradiction to experiment, which shows a
large Mott-Hubbard gap and a prominent pseudogap.  Or instead, one can attempt to dress interaction 
vertices in such a way as to satisfy certain self-consistency conditions \cite{amt}.  More commonly, one
can try a large $N$ expansion, where $N$ represents
the number of degrees of freedom \cite{chubukov}.  Unfortunately, it has
been recently realized that this $1/N$ expansion breaks down at third order for the nematic \cite{max1}
and antiferromagnetic spin fluctuation \cite{max2} problems, casting doubt that a well defined perturbation
expansion exists.

Perhaps the most popular approach of late has been dynamical mean field 
theory (DMFT) \cite{DMFT}.  This is formally based on an expansion in powers of $1/d$ (where $d$ is
the dimensionality) and is related to earlier attempts to understand the physics of $^3$He \cite{vollhardt}.
The most common variants are to expand about the infinite $d$ limit in either real space or momentum
space.  Although convergence can be demonstrated in several model cases, it is not known at this
time how effective the convergence is for models relevant to the cuprates.  Insight on this will be
gained once larger clusters are studied.  Certainly, this approach
has given us insights into the nature of correlations in the cuprates, including the existence of a pseudogap,
and interestingly those clusters which emphasize singlet formation look reminiscent of
the RVB theory of Anderson \cite{kotliar}.

This brings us to more exact techniques.  Obviously, quantum Monte Carlo would in some sense provide
a `solution' to the problem, assuming that the starting point for such simulations, typically the single band 
Hubbard model, is correct (a point which is not generally accepted!).  But this has turned out to be a truly 
difficult undertaking due to the infamous sign problem that occurs when simulating fermions, and which
restricts one from going to too low a temperature.  One can avoid
this by invoking fixed node approximations, or starting with a variational wave function, but these by
definition introduce bias in the simulation.  Such simulations have led to contradictory
results.  That is, at this point in time, we do not know definitively whether the single band Hubbard model
is or is not superconducting \cite{sorella}.

An alternate to quantum Monte Carlo is the density matrix renormalization group (DMRG) 
technique \cite{DMRG}.  This
is in some sense exact in one dimension, and has given some insights as well into two dimensions
(by simulating strips).  For instance, DMRG studies find a tendency towards stripe formation \cite{white}.
Recently, other real space renormalization group techniques have been developed specifically for
two dimensions, including PEPS
(projected entangled pair states) \cite{peps} and MERA (multiscale entanglement renormalization 
ansatz) \cite{mera} which attempt to preserve certain correlations during coarse graining, but the
general efficacy of these methods for the cuprate problem has yet to be demonstrated.

Ultimately, the goal is to determine the source of pairing, but even this has been the source of much
controversy.  For instance, in 1987, spin fluctuation theories predicted the existence of $d_{x^2-y^2}$
symmetry for the pairs \cite{bickers}, but just as in the $^3$He case, this did not prove to be definitive,
despite claims one might see in the popular literature \cite{mann}.
Within a year, RVB theories also found
this ground state.  Moreover, in both cases, the functional form predicted, $\cos(k_xa)-\cos(k_ya)$,
turned out to be in remarkable agreement with what was later observed by photoemission \cite{ding}.
Anderson has suggested that a way to differentiate might be in the frequency dependence of the
pair interaction \cite{mammoth}, in that RVB theories assume an `instantaneous'  pair interaction
(the superexchange $J$ which does not develop dynamics until energies of order the Coulomb
repulsion $U$), whereas spin fluctuation theories have their dynamics set by the scale of the spin
fluctuations (which is order $J$ itself).
There are two issues with this, though.  First, both contributions can be present, though
typically the `instantaneous' contribution is the sub-dominant one \cite{maier}.  Even in RVB theories, the
claim of an instantaneous nature does not quite ring true, since attempts to go beyond the `mean field'
RVB approach typically invoke gauge fluctuations (referring to the constraints involved with no double
occupancy) which introduce significant low energy dynamics \cite{RMP-Patrick}.

An alternate to all of the above is to look directly at the question of the condensation energy, that is, 
where the
energy savings is coming from in forming the superconducting state \cite{tony}.  This has a long history
predating BCS theory, and the answer to this question depends on what energy scale one is looking at.
For instance, in classic superconductors with an isotope coefficient of 1/2, it can be easily demonstrated
that the entire energy savings is coming from the ion kinetic energy \cite{chester}.  In BCS theory, though,
one
projects to a low energy subspace where this effect is absorbed into the definition of the electron potential
energy (i.e., the ions are integrated out).  The net condensation energy comes from a near cancellation 
between a lowering of the potential energy and a raising
of the kinetic energy \cite{tinkham}, the latter being due to particle-hole mixing.  Surprisingly, this simple
picture might not apply to the cuprates.  Evidence from both infrared conductivity \cite{vdM} and
photoemission \cite{cond} indicate that at least for underdoped cuprates, the kinetic energy is actually
lowered below $T_c$.  This occurs because superconductivity is a coherent state which emerges
over much of the doping range
from an incoherent normal state.  The energy savings due to coherency is such that it can overwhelm the loss of
kinetic energy due to particle-hole mixing, leading to a net gain.  Of course, one expects the potential
energy to be lowered as well.  Studies of inelastic neutron scattering (INS) data \cite{dai} find that the
exchange energy is lowered below $T_c$, as expected in spin fluctuation based pictures.  In all cases
where lowering of the energy has been detected, the values determined are far in excess of the actual
condensation energy.  So, either the results are incomplete (for instance, only a small range of momentum
and energy have been used in the INS analysis), or other equally large `energy raising' contributions are
lurking around.

We conclude with a discussion of the phase diagram.  Early on, RVB theories predicted a rather novel
phase diagram, consisting of crossing lines as a function of doping \cite{RMP-Patrick}.
Below a temperature $T^*$ which decreases linearly with
doping, a `spin gap' phase was predicted due to d-wave spin singlet formation.  Below a temperature 
$T_{coh}$ which
increases linearly with doping, the doped charge degrees of freedom become phase coherent.  Only below
both lines, which cross at optimal doping, does one get superconductivity, whereas above both lines, one
has a `strange metal'.  A similar phase diagram,
where $T_{coh}$ is instead the phase stiffness temperature of the pairs, was proposed by Emery 
and Kivelson \cite{Nat95}.  Interestingly, a recent photoemission study claims to see such a phase
diagram \cite{Utpal11}.

This can be contrasted with a `quantum critical' phase diagram \cite{qc}, where a long range (or quasi-long
range) ordered phase is suppressed to zero with doping.  Its `mirror' phase at higher dopings is a
quantum disordered analogue, typically a Fermi liquid.  The former defines $T^*$ and the latter $T_{coh}$
which instead of crossing (like in the RVB scenario) touch at zero temperature near optimal doping.
In this case, the superconducting dome `screens' this singularity, being generated by pairing due to
critical fluctuations associated with the ordered phase.  Above these lines, one has a 
quantum critical
phase (analogue of the `strange metal') characterized by a linear $T$ resistivity.  A variety of experimental
data, including ironically photoemission, have given support for such a phase diagram.

Though both scenarios seem radically different, they do have a point in common in that the nature of the 
pseudogap phase below $T^*$ determines the origin of the superconducting state.  In the RVB approach,
the superconducting state is simply a charged version of the singlet `spin gap' phase.
In the quantum critical approach, the pairing is mediated by critical fluctuations associated with
the pseudogap phase.  Therefore the nature of the pseudogap phase is key.  A variety of
phenomena are associated with this phase, including evidence for nematic distortions (where x-y symmetry
is spontaneously broken), stripes (both charge and spin varieties), and orbital currents.  Many of these
effects are quite subtle, yet the pseudogap itself is very large, with a magnitude that strongly increases with
underdoping.  The jury is still out on its origin, although the present author has his bet on spin singlet
formation.  Regardless, magnetic correlations definitely play a prominent role in the entire doping
range superconductivity is observed \cite{keimer}.  Whether this means RVB, spin
fluctuations, orbital currents, or some combination thereof, such magnetic correlations are the likely 
source of d-wave
pairing.  But building a rigorous strong coupling theory has certainly proven to be a challenge.  Perhaps ideas
from string theory and black hole physics will help in this regard \cite{ads}.  But then again, perhaps not!

\begin{acknowledgements}
This work was supported by the US DOE, Office of Science, under contract 
DE-AC02-06CH11357 and by the Center for Emergent Superconductivity, an Energy Frontier Research
Center funded by the US DOE, Office of Science, under Award No.~DE-AC0298CH1088.
\end{acknowledgements}


\begin{thebibliography}{}

\bibitem{layzer} D. Fay and A. Layzer, Phys. Rev. Lett. {\bf 20}, 187 (1968).

\bibitem{brinkman} P. W. Anderson and W. F. Brinkman, Phys. Rev. Lett. {\bf 30}, 1108 (1973).

\bibitem{leggett} A. J. Leggett, Rev. Mod. Phys. {\bf 47}, 331 (1975).

\bibitem{varma} C. M. Varma, Phys. Rev. B {\bf 73}, 115113 (2006).

\bibitem{pwa} P. W. Anderson, Science {\bf 235}, 1196 (1987).

\bibitem{RMP-Steve} S. A. Kivelson {\it et al.}, Rev. Mod. Phys. {\bf 75}, 1201 (2003).

\bibitem{STM} K. M. Lang {\it et al.}, Nature {\bf 415}, 412 (2002).

\bibitem{RMP-ZX} A. Damascelli, Z. Hussain and Z.-X. Shen, Rev. Mod. Phys. {\bf 75}, 473 (2003).

\bibitem{bcs} J. Bardeen, L. N. Cooper and J. R. Schrieffer, Phys. Rev. {\bf 108}, 1175 (1957).

\bibitem{scalapino} J. R. Schrieffer, D. J. Scalapino and J. W. Wilkins, Phys. Rev. Lett. {\bf 10}, 336 (1963).

\bibitem{AGD} A. A. Abrikosov, L. P. GorÕkov and I. E. Dzyaloshinsky, {\it Methods of Quantum Field Theory 
in Statistical Mechanics} (Dover, New York, 1975).

\bibitem{morel} P. Morel and P. W. Anderson, Phys. Rev. {\bf 125}, 1263 (1962).

\bibitem{cohen} M. Cohen and P. W. Anderson, in {\it Superconductivity in d- and f-band Metals},
ed. D. H. Douglas (AIP, New York, 1972).

\bibitem{chubukov} A. Abanov, A. V. Chubukov and J. Schmalian, Adv. Phys. {\bf 52}, 119 (2003).

\bibitem{levin} J. A. Hertz, K. Levin, M. T. Beal-Monod, Solid State Comm. {\bf 18}, 803 (1976).

\bibitem{flex} N. E. Bickers, D. J. Scalapino and S. R. White, Phys. Rev. Lett. {\bf 62}, 961 (1989).

\bibitem{vilk} Y. M. Vilk and A.-M. S. Tremblay, J. Phys. I {\bf 7}, 1309 (1997).

\bibitem{amt} A.-M. S. Tremblay, B. Kyung and D. Senechal, Low Temp. Phys. {\bf 32}, 424 (2006).

\bibitem{max1} M. A. Metlitski and S. Sachdev, Phys. Rev. B {\bf 82}, 075127 (2010).

\bibitem{max2} M. A. Metlitski and S. Sachdev, Phys. Rev. B {\bf 82}, 075128 (2010).

\bibitem{DMFT} A. Georges, G. Kotliar, W. Krauth and M. J. Rozenberg, Rev. Mod. Phys. {\bf 68}, 13 (1996).

\bibitem{vollhardt} D. Vollhardt, Rev. Mod. Phys. {\bf 56}, 99 (1984).

\bibitem{kotliar} K. Haule and G. Kotliar, Phys. Rev. B {\bf 76}, 104509 (2007).

\bibitem{sorella} S. Sorella {\it et al.}, Phys. Rev. Lett. {\bf 88}, 117002 (2002).

\bibitem{DMRG} S. R. White, Phys. Rev. Lett. {\bf 69}, 2863 (1992).

\bibitem{white} S. R. White and D. J. Scalapino, Phys. Rev. Lett. {\bf 80}, 1272 (1998).

\bibitem{peps} P. Corboz, S. R. White, G. Vidal and M. Troyer, Phys. Rev. B {\bf 84}, 041108(R) (2011).

\bibitem{mera} G. Evenbly and G. Vidal, Phys. Rev. Lett. {\bf 104}, 187203 (2010).

\bibitem{bickers} N. E. Bickers, D. J. Scalapino and R. T. Scaletar, Intl. J. Mod. Phys. B {\bf 1}, 687 (1987).

\bibitem{mann} A. Mann, Nature {\bf 475}, 280 (2011).

\bibitem{ding} H. Ding {\it et al.}, Phys. Rev. B {\bf 54}, R9678 (1996).

\bibitem{mammoth} P. W. Anderson, Science {\bf 317}, 1705 (2007).

\bibitem{maier} T. A. Maier, D. Poilblanc and D. J. Scalapino, Phys. Rev. Lett. {\bf 100}, 237001 (2008).

\bibitem{RMP-Patrick} P. A. Lee, N. Nagaosa and X.-G. Wen, Rev. Mod. Phys. {\bf 78}, 17 (2006).

\bibitem{tony} A. J. Leggett, Nat. Phys. {\bf 2}, 134 (2006).

\bibitem{chester} G. V. Chester, Phys. Rev. {\bf 103}, 1693 (1956).

\bibitem{tinkham} M. Tinkham, {\it Introduction to Superconductivity} (McGraw-Hill, New York, 1966).

\bibitem{vdM} H. J. A. Molegraaf {\it et al.}, Science {\bf 295}, 2239 (2002). 

\bibitem{cond} M. R. Norman, M. Randeria, B. Janko and J. C. Campuzano, Phys. Rev. B {\bf 61}, 14742 (2000).

\bibitem{dai} H. Woo {\it et al.}, Nat. Phys. {\bf 2}, 600 (2006).

\bibitem{Nat95} V. Emery and S. A. Kivelson, Nature {\bf 374}, 434 (1995).

\bibitem{Utpal11} U. Chatterjee {\it et al.}, Proc. Natl. Acad. Sci. {\bf 108}, 9346 (2011).

\bibitem{qc} S. Sachdev, Phys. Status Solidi B {\bf 247}, 537 (2010).

\bibitem{keimer} M. Le Tacon {\it et al.}, Nat. Phys. online (2011).

\bibitem{ads} T. Faulkner {\it et al.}, Science {\bf 329}, 1043 (2010).

\end{thebibliography}
\end{document}